%% file: main.tex
\pdfoutput=1
\documentclass{llncs}
\usepackage{llncsdoc}
\usepackage[letterpaper,margin=1in]{geometry}
\usepackage{amsmath, amssymb, amsfonts}
\usepackage{ifthen}
\usepackage{arrayjobx}
\usepackage{tikz}
\usetikzlibrary{positioning,decorations.pathreplacing}

\usepackage{cite}
\usepackage{appendix}
\usepackage{graphicx}
\usepackage{color}
\usepackage[noend]{algpseudocode}
\usepackage{epstopdf}
\usepackage{wrapfig}
\usepackage[framemethod=tikz]{mdframed}
\usepackage[bottom]{footmisc}
\usepackage{enumitem}
\setitemize{noitemsep,topsep=0pt,parsep=0pt,partopsep=0pt}
\usepackage{caption}
\usepackage{xspace}
\usepackage{hyperref}
\hypersetup{
    unicode=false,          
    colorlinks=true,        
    linkcolor=red,          
    citecolor=green,        
    filecolor=magenta,      
    urlcolor=cyan           
}
\usepackage[capitalize]{cleveref}
\usepackage{fancyhdr} 
\usepackage{color} 
\usepackage[nofillcomment,linesnumbered,noend,noresetcount,noline]{algorithm2e}
\usepackage{graphicx} 
\usepackage{complexity} 
\usepackage{enumitem} 
\usepackage{shorttoc} 
\usepackage{array} 
\usepackage{float}
\usepackage{pdfsync}
\usepackage{verbatim}
\hypersetup{pageanchor=false}

\AtBeginDocument{%
 \abovedisplayskip=2pt plus 1pt minus 1pt
 \abovedisplayshortskip=0pt plus 1pt
 \belowdisplayskip=2pt plus 1pt minus 1pt
 \belowdisplayshortskip=0pt plus 1pt
}

\algnewcommand\algorithmicswitch{\textbf{switch}}
\algnewcommand\algorithmiccase{\textbf{case}}

\algdef{SE}[SWITCH]{Switch}{EndSwitch}[1]{\algorithmicswitch\ #1\ \algorithmicdo}{\algorithmicend\ \algorithmicswitch}%
\algdef{SE}[CASE]{Case}{EndCase}[1]{\algorithmiccase\ #1}{\algorithmicend\ \algorithmiccase}%
\algtext*{EndSwitch}%
\algtext*{EndCase}%

\renewcommand{\paragraph}[1]{\vspace{0.15cm}\noindent {\bf #1}:}

\newcommand{\idlow}[1]{\mathord{\mathcode`\-="702D\it #1\mathcode`\-="2200}}
\newcommand{\id}[1]{\ensuremath{\idlow{#1}}}
\newcommand{\litlow}[1]{\mathord{\mathcode`\-="702D\sf #1\mathcode`\-="2200}}
\newcommand{\lit}[1]{\ensuremath{\litlow{#1}}}

\newcommand{\namedref}[2]{\hyperref[#2]{#1~\ref*{#2}}}

\newcommand{\theoremref}[1]{\namedref{Theorem}{#1}}

\newcommand{\figureref}[1]{\namedref{Figure}{#1}}
\newcommand{\figurerefb}[2]{\hyperref[#1]{Figure~\ref*{#1}#2}}

\newcommand{\lemmaref}[1]{\namedref{Lemma}{#1}}

\newcommand{\propositionref}[1]{\namedref{Proposition}{#1}}

\newcommand{\equationref}[1]{\hyperref[#1]{(\ref*{#1})}}
\renewcommand{\eqref}{\equationref}

\usepackage[textsize=tiny]{todonotes}

\newcommand{\DEBUG}[1]{}

\newcommand{\FullOrShort}{short}

\ifthenelse{\equal{\FullOrShort}{full}}{
        
  \newcommand{\fullOnly}[1]{#1}
  \newcommand{\shortOnly}[1]{}
  
  }{

    \newcommand{\fullOnly}[1]{}
    \newcommand{\shortOnly}[1]{#1}
    
  }

\begin{document}

\date{}

\title{On the Importance of Registers for Computability}

\author{
  Rati Gelashvili  
  \and
  Mohsen Ghaffari 
  \and
  Jerry Li 
  \and
  Nir Shavit 
}

\institute{MIT \\ {\small \email{\{gelash, ghaffari, jerryzli\}@mit.edu; shanir@csail.mit.edu}}}

\maketitle
\begin{abstract}
All consensus hierarchies in the literature assume that we have, 
        in addition to copies of a given object, an unbounded number of registers. 
But why do we really need these registers?

This paper considers what would happen if one attempts to solve consensus using various objects 
        but without any registers. 
We show that under a reasonable assumption, objects like queues and stacks cannot emulate 
        the missing registers. 
We also show that, perhaps surprisingly, initialization, shown to have no computational consequences
        when registers are readily available, is crucial 
        in determining the synchronization power of objects when no registers are allowed. 
Finally, we show that without registers, the number of available objects 
        affects the level of consensus that can be solved.

Our work thus raises the question of whether consensus hierarchies which assume an unbounded number 
        of registers truly capture synchronization power, and begins a line of research aimed at 
        better understanding the interaction between read-write memory and 
        the powerful synchronization operations available on modern architectures.
\end{abstract}

\input{intro.tex}
\input{collect.tex}
\input{imposs.tex}

\section{Acknowledgements}
Support is gratefully acknowledged from the National Science Foundation under grants CCF-1217921, 
        CCF-1301926, and IIS-1447786, the Department of Energy under grant ER26116/DE-SC0008923, 
        and the Oracle and Intel corporations.
        
The authors would like to thank Eli Gafni and Yehuda Afek for helpful conversations and feedback.

\bibliographystyle{alpha}
\footnotesize
\bibliography{biblio}
        

\end{document}

%% file: intro.tex
\section{Introduction}
In a seminal paper~\cite{herlihy1991}, Herlihy introduced the \emph{consensus hierarchy}, 
        where the synchronization power of an object is measured by its \emph{consensus number}, 
        defined as the maximum number of processes for which wait-free consensus is solvable using 
        instances of the object and as many \textit{read-write registers} as needed.
But do we really need these read-write registers?
In this paper we consider what would happen if one attempts to solve consensus 
        (henceforth we will use the term "solve" to mean a wait-free solution) 
        using various objects without any registers. 
                
Consider the following interesting example. 
It is well known~\cite{herlihy1991} that a single queue initialized with two items
        and with two registers, can solve two process consensus.
We show that this is possible even if the queue is in an arbitrary initial state,
        and that a queue can solve two process consensus even without registers 
        if it is initialized properly.
Moreover, two queues in arbitrary initial states are sufficient for solving two process consensus. 
On the other hand, we prove that it is impossible to solve  
        two process consensus using a single empty queue.
In other words, unless you have multiple queues or multiple registers, a queue's ability to solve 
        consensus is completely dependent on its initialization. 
This example motivates us to better understand the computational effects of the number 
        of objects and their initialization when no registers are available.  

We begin our investigation by considering a general class of objects we refer to 
        as \emph{consistent sets}, that includes natural objects such as queues, 
        stacks and priority queues.
Most of the above examples for queues are specific instances of our results 
        for consistent set objects.
We show that it is possible to solve two process consensus with a single 
        consistent set object and two registers or with two consistent set objects, 
        even when the objects are initialized in arbitrary states.
We also show the corresponding generalization for the impossibility result mentioned above:

\vspace{6pt}
\noindent \textbf{Theorem 1.}
\textit{
It is impossible to solve consensus for two processes
        using a single consistent set object initialized in an empty state.
}
\vspace{6pt}

\noindent As far as we know this is the first result showing that 
        initialization to a different natural state matters for reaching agreement.
At its core, the proof involves inductively constructing an interleaving of two solitary executions,
        such that the processes cannot distinguish between running alone and running in this
        interleaved execution.
However, obtaining the indistinguishability guarantees is rather involved.
It requires a new technique to adapt the interleaving to the state of the consistent set object, 
        and involves constructing successive pieces of the interleaved execution separately and 
        then merging them.
The challenge is to maintain indistinguishability, which we prove is possible because of the 
        properties of a consistent set object.
        
We have so far focused on whether two processes can solve consensus using 
        a limited number of objects. 
This question has practical value as typically small numbers of objects are used in most 
        data structure implementations. 
However, on the more theoretical side, the work of Jayanti~\cite{jayanti97} shows that robust 
        consensus hierarchies must allow an arbitrary number of objects.
Here we will assume that processes communicate using an unlimited supply of linearizable 
        objects~\cite{HW90}, and as in~\cite{GMT2001,MT2000,ABDPR90}, we will also assume 
        that there are an unlimited number of processes in the system.
Although, in this setting, our impossibility results will still hold in a weaker model where 
        only a bounded number of processes are allowed to run concurrently.
(In fact, even if the algorithms can assume that only two processes will ever run at the same time). 

Let us say that an implementation is \emph{isolation-bounded} if the following holds:
        there exists an absolute constant $M$, such that when the very first method call is executed 
        in complete isolation, it takes at most $M$ steps.
Practically all natural algorithms are isolation-bounded,
        even when an unbounded number of processes are allowed to be concurrent.
For example, all algorithms where the step-complexity of a method can be upper-bounded by a function 
        of the maximum contention (number of concurrent processes) encountered are isolation-bounded.
We will henceforth consider isolation-bounded implementations. 

Consider the test-and-set task~\cite{AGTV92}, a simplification of consensus in which 
        exactly one process knows it is the winner (returns 1) and all other processes know that 
        they are losers (return 0), and assume a corresponding linearizable test-and-set object. 

We begin by showing the following results that capture the effects of having registers:
                
\vspace{6pt}
\noindent \textbf{Theorem 2.}
\textit{
It is impossible to implement an isolation-bounded test-and-set object
        for an unbounded number of processes
        using any number of (possibly infinitely many) empty queues (or empty stacks).
}
\vspace{6pt}
        
\noindent The proof of this theorem is interesting as it follows along lines that have, 
        as far as we know, never been used before in deriving shared-memory lower bounds. 
Essentially, we wish to reduce the general case in which infinitely many processes access infinitely 
        many queues, to the case where infinitely many processes access only finitely many queues 
        in their solo executions.
Once reduced, we can use a counting argument to find two processes whose solo executions can be 
        interleaved so that for both processes running in the interleaved execution, 
        their execution is indistinguishable from running alone. 
To achieve this reduction, we use an argument, akin to diagonalization, to produce an infinite 
        set of processes for which the desired property essentially holds.\footnote{We remark that 
        our proof requires the axiom of countable choice, 
        which we will assume without comment when necessary.}

On the other hand, if read-write registers are available, 
        one can use the tournament tree construction from~\cite{AAGGG10} to get the following result

\vspace{6pt}
\noindent \textbf{Theorem 3.}
\textit{
There is an implementation of an isolation-bounded test-and-set object for an unbounded number of 
        processes using infinitely many consistent set objects 
        (in any initial configuration) and read-write registers.
}
\vspace{6pt}

These theorems have a few important corollaries.
The first of these corollaries demonstrates a fundamental difference between registers 
        and objects like stacks and queues. 

\vspace{6pt}
\noindent \textbf{Corollary 1.}
\textit{
It is impossible to implement a read-write register in an isolation-bounded way
        using any number of (possibly infinitely many) empty queues (stacks).
}
\vspace{6pt}

\noindent Interestingly, if number of processors in the system is bounded, 
        simulations a read-write register exist~\cite{BNP97}.
        
The second corollary is about initialization.
Algorithms for consensus usually assume that the objects and registers are 
        initialized in a certain way.
In fact, the consensus number of an object can change depending on the initial state.
Consider an object with a consensus number at least two
        that has an additional ``invalid" state, unreachable from all other states,
        such that in the invalid state, all method calls return $\id{null}$.
Clearly, the object initialized in the invalid state has consensus number one.

But generally, in most initial states the object will have the same consensus number.
For instance, as shown in~\cite{ABG1994}, this is always true for states reachable 
        from each other.\footnote{In the above example where the consensus number changed, 
        no state was reachable from the invalid state.}
Our second corollary shows that perhaps surprisingly, for some objects the difference in the 
        synchronization power in these initial states can still be quite significant:

\vspace{6pt}
\noindent \textbf{Corollary 2.}
\textit{
It is impossible to implement a queue (a stack) containing one element in its initial state
        using any number of (possibly infinitely many) empty queues (stacks) 
        in an isolation-bounded way.}

%% file: collect.tex
\section{Consistent Sets and Two Consensus}
Let us define a class of objects, that we will call \emph{consistent sets}.
Each consistent set object represents a data-structure of items 
        and implements two linearizable methods:
        \lit{insert(\id{item})} and \lit{remove()}.
We say that a consistent set object \emph{contains} an item, if the item has not been removed since 
        its last insertion in the set.
Assume that $s_1, s_2, \ldots, s_m$ are the items contained in some consistent set object, 
        whereby $s_1$ was inserted before $s_2$, etc, before $s_m$.
The \lit{remove()} operation returns one of the items $s_i$, selected based on a fixed function $F$,
        i.e. $s_i = F(s_1, s_2, \ldots, s_m)$.
If $m=0$, then a special value $\id{null}$ (which can never be an item contained in the set) 
        is returned instead.
A consistent set object can be initiliazed to an empty state (containing $0$ items), 
        or with any finite number of items pre-inserted in an arbitrary fixed order.

Each consistent set object has its function $F$, defined for all possible item sequences
        that satisfies the following two \emph{consistency} properties:
\begin{itemize}
\item If there exist (possibly empty) sequences of items $L, M, R$, 
        such that $F(L,s_i,M,s_j,R) = s_i$, 
        then there do not exist item sequences (represented by dots), 
        so that $F(\ldots,s_i,\ldots,s_j,\ldots) = s_j$.
\item If there exist (possibly empty) sequences of items $L, M, R$, 
        such that $F(L,s_i,M,s_j,R) = s_j$, 
        then there do not exist possible item sequences (represented by dots), 
        so that $F(\ldots,s_i,\ldots,s_j,\ldots) = s_i$.
\end{itemize}
\noindent The exact choice of function $F$ determines precise semantics of the data-structure.
For instance, a first-in-first-out queue, a stack and a priority queue are all consistent set
        objects and correspond to particular choices of $F$:
        for a queue $F(s_1,\ldots,s_m) = s_1$, for stack $F$ picks $s_m$ and 
        for a priority queue it picks the item with the maximum (minimum) priority.
\begin{lemma}
\label{lem:collcons2}
It is possible to solve wait-free two process consensus using any consistent set object $\id{O}$, 
        initialized with a finite number of arbitrary items in an arbitrary order.
\end{lemma}
\begin{proof}
Let $\id{W}$ be an item that is different from all initial items in $\id{O}$. 
We claim that the algorithm described in pseudo-code on~\figureref{fig:collcons2} solves 
        wait-free consensus for two processes. 
It is straightforward to show wait-freedom, so it suffices to demonstrate that 
        the algorithm solves consensus. 
It is also straightforward to show that each process returns either its own value 
        or the other process's value. 
For $i \in \{0, 1\}$, let $v_i$ denote the value that process $i$ gets as input. 
Suppose for the sake of contradiction that the processes return different values. 
There are two cases.

\paragraph{Process $i$ returns $v_i$, for $i \in \{0, 1\}$} 
By inspection, the only way that process $1$ can return $v_1$ is if it returns at line $9$, that is,
        it enters the while loop then removes $W$. 
There are two sub-cases. 
Suppose process 0 returns on line 4, so that it returned since it saw $\id{Proposed}[1] = \bot$, 
        and returns $v_0$. 
By inspection, this is only possible if this occurs before process 1 executes line 3, which implies 
        that process 0 executes line 2 before process 1 executes line 4, 
        which implies that when process 1 reads $\id{Proposed}[0]$ on line 4, it will see $v_0$, 
        and thus will return it, which is a contradiction. 
Alternatively, process $0$ could return on line $8$, but this would imply that on line $7$, in some 
        iteration of the loop, removes $W$. Since $W$ is only inserted once into the consistent set,
        this is a contradiction, since process 1 must remove it as well.

\paragraph{Process $i$ returns $v_{1 - i}$, for $i \in \{0, 1\}$} 
By inspection, the only way that process $0$ can return $v_1$ is if it returns on line 10, that is, 
        it sees an empty consistent set. 
There are again two sub-cases, since process 1 can return $v_0$ in one of two ways. 
Suppose process 1 returns on line 5. 
Then by that point in the execution, process 1 has already executed $\id{O}.\lit{insert(W)}$. 
Then, when process 0 enters the while loop, it is guaranteed to eventually remove $W$ since it is 
        the only process removing elements from the consistent set, so it will return $v_0$ as well,
        which is a contradiction. 
Thus, suppose process 1 returns on line 11. 
But this happens after process 1 performs $\id{O}.\lit{insert}(W)$, and neither process can see $W$ 
        while removing elements from the consistent set until the set is empty, 
        which is a contradiction.
\input{algowithregs.tex}
\end{proof}
\noindent Let us next consider the synchronization power of consistent sets without registers.
\begin{lemma}
\label{lem:2obj}
It is possible to solve wait-free two process consensus 
        using any two consistent set objects $\id{O_0}$ and $\id{O_1}$, 
        initialized with a finite number of arbitrary items in an arbitrary order.
\end{lemma}
\begin{proof}
The algorithm is described on~\figureref{fig:2objcons2}. 
Recall $F$ is the function which uniquely defines the consistent set.
We have two consistent set objects: 
        $\id{O_0}$, where process $\id{O}$ inserts to, 
        and $\id{O_1}$, where process $1$ inserts to.
Inserted elements are pairs of form $\{\id{P_i},v_i\}$ and $\{\id{Q_i},v_i\}$, 
        where $v_i$ is the input of process $i$, and $\id{P_i}$ or $\id{Q_i}$ are two 
        different prefixes, such that the corresponding pairs are not the same 
        as any of the initial items in sets $\id{O_i}$.
        
We claim that the algorithm solves consensus. 
As with the proof of~\lemmaref{lem:collcons2}, let $v_i$ be the input of the process $i$, 
        for $i \in \{0, 1\}$. 
It is again straightforward to see that the algorithm is wait-free. 
Thus it suffices to prove that the processes will return the same value. 
Suppose for the sake of contradiction that the processes return different values. 
Notice by the definition of a consistent set, if a process's call to $\lit{remLW}(\id{O})$ returns 
        $\{L, v\}$, then there must have been a previous $\lit{remove}$ operation performed on 
        $\id{O}$ which returned the unique other element $e$ inserted into $\id{O}$ with 
        $e.\lit{second} = v$ and $e.\lit{first} \in \{P_0, P_1, Q_0, Q_1\}$.
Moreover, if $e$ was removed due to a $\lit{remLW}$ operation, that operation would 
        return $\{W, v\}$. 

There are two cases.

\paragraph{Process $i$ returns $v_i$, for $i \in \{0, 1\}$} 
By inspection, there is one way for process 0 to return $v_0$, which is to return on line 7, 
        which implies that $a_0.\lit{first} = W$ and $a_1 = \id{null}$. 
That $a_1 = \id{null}$ implies that process 0 executes line 4 before process 1 executes line 13, 
        which implies that $b_0 \neq \id{null}$. 
Since $a_0.\lit{first} = W$, this implies that $b_0.\lit{first} = L$. 
Moreover, since $a_1 = \id{null}$, this implies that $b_1.\lit{first} = W$, 
        which is a contradiction, as then process 1 cannot return $v_1$.

\paragraph{Process $i$ returns $v_{1 - i}$, for $i \in \{0, 1\}$} 
By inspection, there is one way for process 1 to return $v_0$, which is for it to fail the 
        if statement on line 17. 
To fail this if statement means that $b_0.\lit{first} = L$ and $b_1.\lit{first} = W$ 
        (since $b_1 \neq \id{null}$). 
Since $b_0.\lit{first} = L$, this implies that process 1 finishes line 15 after process 1 
        finishes line 5, and it also implies that $a_0.\lit{first} = W$. 
This implies that process $0$ finishes executing line 4 before process 1 starts executing line 16, 
        so the only way that $b_1.\lit{first} = W$ is if $a_1 = \id{null}$, 
        thus process $0$ will return $v_0$ as well. 
\input{algo2obj.tex}
\end{proof}
\noindent Any algorithm for two-consensus (including the algorithms above) can be used to solve 
        test-and-set for two processes, 
        simply by having each process return $1$ instead of its own value and $0$ otherwise.
        
Let us call a state of an instance of any consistent set object $\id{O}$ \emph{lucky}, 
        if it contains only a single copy of some item $\id{W}$.
\begin{lemma}
\label{lem:collinftas}
It is possible to implement a test-and-set object for an unbounded number of processes using 
        a single consistent set object $\id{O}$ initialized in a lucky state.
\end{lemma}
\begin{proof}
The algorithm for each process is to simply remove items from $\id{O}$ 
        until observing $\id{W}$ or $\id{null}$.
In the first case, the process returns $1$ and in the second case, it returns $0$.
By the semantics of the data-structure, one and only one process will remove $\id{W}$ 
        and return $1$.
Moreover, that process can in fact be linearized as the winner of the test-and-set, i.e. 
        as the first to call the $\lit{test-and-set}()$ method 
        (since otherwise, another method call must have completed strictly earlier and 
        that it would have removed the unique element $\id{W}$).   
\end{proof}
\begin{lemma}
\label{lem:queue2cons}
There exists a consistent set object $\id{O}$, 
        such that it is possible to solve wait-free two process consensus  
        with $\id{O}$ initialized in a lucky state.
\end{lemma}
\begin{proof}
A first-in-first-out queue is such an object. 
The algorithm for each process is to first enqueue its own item and then keep dequeuing until 
        either observing $\id{W}$ or $\id{null}$. 
In the first case, the process returns own value.
Otherwise, it returns the value of the other process (we show below how), and the exact argument 
        from~\lemmaref{lem:collinftas} finishes the correctness proof. 
        
To show how the process knows the value to return, 
        consider the process $p$ that observes $\id{null}$ at time $t$.
Since the other process has dequeued $\id{W}$ by time $t$, 
        it must have already enqueued its value, which comes later than 
        all original items of $O$ (including $\id{W}$) in the first-in-first-out order.
The other item with this property is the input value of $p$ itself.
Therefore, the last two items dequeued by $p$ must be the input values of the processes,
        $p$ knows its own value and can simply tell the value of the other process.
\end{proof}
\noindent Given these insights, the following result may be surprising:
\begin{theorem}
\label{thm:emptyno}
It is impossible to solve wait-free two process consensus
        using a single consistent set object $\id{O}$ initialized in an empty state.
\end{theorem}
\begin{proof}
Assume the contrary.
Then the existence of the consensus protocol implies that 
        there also exists a wait-free test-and-set implementation for two processes using just 
        a single consistent set object $\id{O}$ initialized in an empty state.
For each process $i \in \{0, 1\}$ there exists a solo execution where process $i$ runs in isolation 
        and returns $1$ after some finite number $t_i$ of steps.
Let $E_0$ and $E_1$ be these solo executions.
Each step in these executions is either an \lit{insert(\id{item})} or \lit{remove()} 
        call on $\id{O}$.

We obtain a contradiction by constructing a schedule where both processes are executed, 
        but never observe any difference from their solo executions, 
        i.e. the execution of process $i$ is indistinguishable from $E_i$ from its prospective.
Formally, given a serial execution $E_i$ which only makes method calls to $O$, and a linearized execution $E$ containing $E_i$ and      
        other method calls from other processes to $O$, we say that $E_i$ is \emph{indistinguishable} from $E$ if 
        for every remove operation in $E_i$, it gets the same response as it does in $E$. 
Clearly, if process $i$ has solo execution $E_i$ and $E$ is an execution which is indistinguishable from $E_i$, it must return $1$ in $E$, so if an execution $E$ is indistinguishable from two solo executions, we derive a contradiction. 

To construct this interleaving, we use induction on total number of steps in $E_0$ and $E_1$ to prove the existence of 
        the interleaved execution. We say the first $\ell$ steps of an execution form an \emph{$\ell$-prefix}.

The following proposition provides the base case for induction.
\begin{proposition}
If for one of the processes, say for process $j$, $t_j = 0$ holds, then it is possible to interleave
        the executions $E_0$ and $E_1$ such that the interleaved execution is indistinguishable from
        the solo execution for each process.
\end{proposition}
\begin{proof}
The number of steps in solo execution $E_j$ is $0$, 
        so we start by running process $j$ which immediately returns as in $E_j$ 
        and does not change the state of the object $\id{O}$. 
Thus we then complete the interleaved execution by running process $1-j$ until it returns,
        and because the starting state of $\id{O}$ is empty as in $E_{1-j}$,
        this execution also precisely matches $E_{1-j}$.
\end{proof}

\noindent
For inductive step, assume we know that if the total number of steps in two solo executions $E_0$ and $E_1$ is 
        less than $k$, then it is possible to interleave them such that the interleaved execution is 
        indistinguishable from the solo execution for each process.

We now consider several cases, each requiring a different treatment.
By adjusting formulations it is possible to merge some cases, 
        but the particular structure is chosen for clarity. Let the total number of steps in $E_0$ and $E_1$ be $k$.
        
\paragraph{Case 1: A mute prefix} An $\ell$-prefix for a solo execution for process $i$ is called \emph{mute} if $\id{O}$ remains empty after the prefix 
      is executed by process $i$ in isolation.
        
\begin{proposition}
If one of the executions, say execution $E_j$ contains a non-empty mute prefix, then it is possible 
        to interleave the executions $E_0$ and $E_1$ such that the interleaved execution is 
        indistinguishable from the solo execution for each process.
\end{proposition}
\begin{proof}
We start the interleaved execution by letting process $j$ execute the mute prefix of $E_j$. 
This is possible because we actually run process $j$ in isolation, 
        so it simply executes the mute prefix exactly as in $E_j$.
Afterwards, by definition of the mute prefix, $\id{O}$ is empty.
Moreover, the total number of steps in the solo executions that the rest of the interleaved 
        execution should match has strictly decreased.
Therefore, we can use the inductive hypothesis for the same $E_{1-j}$ and $E_j$ without the 
        non-empty prefix to construct the rest of the interleaved execution.
\end{proof}
  
\noindent Thus we may assume that the solo execution $E_i$ for process $i \in \{0, 1\}$ does not contain a mute prefix and it consists of non-zero number of steps.
Define $f_i(\ell)$ to be the item that would be removed by 
        a \lit{remove()} call right after executing an $\ell$-prefix of $E_i$ in isolation.

\paragraph{Case 2: A barrier} For $i \in \{0, 1\}$, let $s_1, s_2, \ldots, s_m$ be the items that are inserted and removed 
        from $\id{O}$ during the solo execution $E_i$ by process $i$, in order of their insertion.
 Let $g_i$ be the item that would be removed the last if we first inserted all of these items 
        in $\id{O}$ in order, and then removed them one-by-one.
Note that this does not have to be $s_m$.
We call $f_i(\ell)$ a \emph{barrier} if $F(f_i(\ell), g_{1-i}) = g_{1-i}$.

\begin{example}
The motivating example of a barrier is when $\id{O}$ is a priority queue which returns elements with high priority first. Consider the situation where process 0 (say) inserts a number of elements into the priority queue with priority $\leq 1$ then some elements with priority $2$ in its solo execution, and process 1 inserts many elements into $\id{O}$ with priorities either $2$ or $3$ in its solo execution. Then, the prefix of process 0 which consists of it inserting elements with priority $\leq 1$ forms a barrier, and such a prefix is natural to consider because this essentially acts like a mute prefix to process 1 in that process 1 will never see anything from this prefix, and mute prefixes are easy to induct on.
\end{example}

To reason about this case, we need a technical property about the behavior of consistent sets which is obvious
for simple objects such as queues, stacks, and priority queues.

\begin{proposition}
Consider a serial execution $E$ consisting of calls to a consistent set object $O$. Let $s$ be some element
inserted and subsequently removed during $E$, and let $E'$ be the execution constructed by removing $\lit{insert}(s)$
and the $\lit{remove()}$ which returned $s$. Then the output of all other $\lit{remove()}$ operations in $E'$ is unchanged.
\end{proposition}

\begin{proof}
We will actually prove a slightly stronger statement: that at any point in the execution $E$, if $O$ contains $s$, at that same point in time in $E'$, the state of $O$ is identical except with $s$ removed, and if $O$ does not contain $s$. then at the same point in time in $E'$, the state of $O$ is exactly the same. This clearly implies our claim. 

To prove this stronger statement, we proceed by contradiction. Let $R_1$ be the first operation after which the states of $O$ in $E$ and $E'$ do not follow this invariant. By inspection this must be a remove operation. Denote the  $\lit{remove()}$ which returned $s$ by $R$.
Clearly the behavior of $O$ at any state before $\lit{insert}(s)$ occurs is the same in $E$ and $E'$, so $R_1$ must happen after the insertion of $s$. Similarly, if $R_1$ was after $R$ in $E$, then by the invariant, before $R$ the state of $O$ in $E$ and $E'$ is identical.
Thus the last remaining case is if $R_1$ was scheduled before $R$ in $E$ but after $\lit{insert}(s)$. 
Suppose in $E$ it returns some element $s'$ and in $E'$ it returns some element $s'' \neq s'$.
Let $A = s_1, \ldots, s_\ell$ be the list of objects in present in $O$ ordered by insertion time
 if we execute $E$ but pause right before executing $R_1$. 
 Clearly this is of the form $L, s', M, s'', R$ or $L, s'', M, s', R$ for some $L, M, R$, where $s$ is in either $L, M,$ or $R$. W.l.o.g. assume that 
 it is of the former type, and assume $s \in L$ (the other cases are identical). We know that
 $F(A) = s'$. Form $L'$ by removing $s$ from $L$, and let $A' = L', s', M, s'', R$. Then by consistency, $F(A') \neq s''$. But by the invariant, before $R_1$, the state
 of $O$ in $E'$ was exactly $A'$, which is impossible. This proves the proposition. 
\end{proof}

Now we have the tools to do the induction in the presence of a barrier:

\begin{proposition}
\label{prp:barrier}
If one of the executions, say execution $E_j$, contains a barrier $f_j(\ell)$, then it is possible 
        to interleave the executions $E_0$ and $E_1$ such that the interleaved execution is 
        indistinguishable from the solo execution for each process.
\end{proposition}
\begin{proof}
Consider the largest $\ell$ so that the $\ell$-prefix of $E_j$ is a barrier. 
We start building the desired interleaved execution by executing the $\ell$-prefix $p_j$ of $E_j$.
This leaves a number of items in $\id{O}$, so in particular $f_j (\ell)$ is well-defined.
Now, let us \emph{trim} the remaining piece of $E_j$: we get rid of all $\lit{remove()}$ operations 
        that in the solo execution remove items inserted in $p_j$.
Thus, the \emph{trimmed} schedule $\tilde{E}_j$ does not contain the $l$-prefix of $E_j$ and 
        any later $\lit{remove()}$ operations that in the solo execution 
        return items inserted during the $l$-prefix. By the above proposition, every $\lit{remove()}$ operation in $\tilde{E}_j$
        returns the same thing it did in $E_j$. In particular, none of them return $\id{null}$ because none of them could have returned null
        in $E_j$ as otherwise $E_j$ would have had a mute prefix.
        
Because the number of operations in $\tilde{E}_j$ is strictly smaller than in $E_j$, using our 
        inductive hypothesis let us construct an indistinguishable interleaved execution $X$ for 
        executions $\tilde{E}_j$ and $E_{1-j}$ assuming that $\id{O}$ started in an empty state.
Note that execution $X$ is only indistinguishable if $\id{O}$ is initially empty and 
        moreover, it does not immediately provide any guarantees for the original execution $E_j$.

However, we will show that it is possible to interleave the trimmed operations from $E_j$ 
        back into $X$ to create $X'$ so that $p_j X'$ is a valid interleaving of $E_0$ and $E_1$ and 
        is indistinguishable to both processes from their solo executions.
Assume the opposite, and consider first time $t$ at which we are unable to indistinguishably 
        schedule the next operation without violating the above invariant. 
Since the only operations which provide feedback are $\lit{remove()}$ operations, 
        we can assume without the loss of generality that the next operations to be scheduled 
        for both processes are both $\lit{remove()}$ operations.

Suppose at time $t$, the next operation scheduled in $X$ is by process $1-j$.
The operation has to be a $\lit{remove()}$ that returns some item $s$ instead of another item 
        $r \neq s$ that would be returned at this point in $E_{1-j}$.
By our assumption, all previous operations have been indistinguishable, 
        so $\id{O}$ has to contain item $r$ at time $t$.
Also, $r$ is clearly inserted by process $1-j$,
        since it is removed by process $1-j$ in the solo execution $E_{1-j}$.
If $s$ was inserted during $X$ (and not in $p_j$), since we still insert the items according to $X$ 
        in the new interleaved execution, during the corresponding $\lit{remove()}$ operation in $X$
        items $s$ and $r$ would certainly be contained in $\id{O}$ in the exact same order 
        as during the above $\lit{remove()}$ operation in the interleaved execution.
But since $X$ is indistinguishable from $E_{1-j}$,
        the removal in $X$ returns $r$ and not $s$, contradicting the consistency of $\id{O}$.

If $s$ was inserted during $p_j$, let us w.l.o.g. 
        assume that $f_j(\ell)$ was inserted after $s$ and $g_{1-j}$ after $r$. We will show that $F(s, r) = r$, a contradiction
        since that means that the remove operation at time $t$ would return $r$ instead of $s$, as $s$ is inserted before $r$ in the
        execution of interest since it was inserted during $p_j$.
Consider $u = F(s,f_j(\ell),r,g_{1-j})$.\footnote{The other cases are symmetric: we would consider 
        $F(f_j(l),s,r,g_{1-j})$, $F(s,f_j(\ell),g_{1-j},r)$ or $F(f_j(\ell),s,g_{1-j},r)$.}
We know $F(r,g_{1-j})=r$ by the definition of $g_{1-j}$, so $u \neq g_{1 - j}$ by the definition of consistent sets.
Similarly, since $F(f_j(\ell),g_{1-j}) = g_{1-j}$ since $f_j(\ell)$ is a barrier, we know $u \neq f_j(\ell)$.
 Finally, $F(s,f_j(\ell)) = f_j(\ell)$ by definition of $f_j(\ell)$, so we know that $u \neq s$.
Thus, $u = r$, and so by the properties of consistent sets we conclude that $F(s, r) = r$.
        
Now assume that the next operation according to $X$ is by process $j$. The next operation to be scheduled
for $E_j$ must be a remove (which may have been trimmed). Call this operation $R$.
By assumption, it removes 
        some item $s$ instead of an item $r \neq s$ which would be removed in $E_j$ at this step. If $s$ was inserted by process $j$, then in solo execution $E_j$ process $j$ 
        should have observed items $s$ and $r$ in $\id{O}$ in the same order as here, 
        but removed $r$, contradicting the consistency property.  

Thus suppose $s$ was inserted by process $1-j$. We claim that $R$ must have been trimmed, since otherwise
  $R$ is the next remove operation in execution $X$. But then, since all the items present in $\id{O}$ at this point in $X$ 
 must also be present in $\id{O}$ in this point in the execution we are building, since we have included all the actions of $X$
 up to this point in our execution, this implies by the definition of consistent set objects, that in $X$, $R$ must also remove $s$,
         contradicting the indistinguishability of $X$ from solo executions.
       
But if $R$ was trimmed and would at this point return some $s$ inserted by process $1 - j$, we claim that there exists a $\ell' > \ell$
so that the $\ell'$-prefix of $E_j$ would also be a barrier, which contradicts our choice of $\ell$. Indeed, let $r$ be the item that $R$, the last $\lit{remove()}$ up to this point in the solo execution $E_j$, removes and
        let $v$ be the item that would be removed if we executed another $\lit{remove()}$ 
        right after $E_j$ ($v$ has to exist, otherwise the whole execution $E_j$ is a mute prefix).        
Since the removal of $r$ is trimmed, $\lit{insert(r)}$ must be in the $p_j$.
Assume without the loss of generality that $f_j(l)$ is inserted after $r$ and before $v$ in $E_j$
        and consider $F(r,f_j(l),v)$.\footnote{Otherwise, 
        considering the respective order works analogously}
$F(r, f_j(l)) = f_j(l)$ must hold by the definition of $f_j(l)$, and since the last trimmed removal 
        also observed $v$ but removed $r$, $F(r,v) = r$ holds.
By the definition of a barrier, $F(f_j(l),g_{1-j}) = g_{1-j}$, and so combining these three facts and using consistency like before
        we get $F(r,f_j(l),v,g_{1-j}) = g_{1-j}$ which again by consistency of $F$ 
        implies that $F(v, g_{1-j}) = g_{1-j}$. Thus if we take the prefix of $E_j$ up to and including $R$, we get another barrier
        which has length strictly larger than $\ell$, which is a contradiction.
This completes the proof of the proposition.

\end{proof}

\paragraph{Case 3: No mute prefixes or barriers} 
The rest of the proof of the main theorem 
        considers the case when none of the executions $E_i$ ($i \in \{0,1\}$) contains 
        a mute prefix or a barrier.
The application of the inductive hypothesis (albeit twice) and the trimming technique 
        is still required, but the partitioning of executions and the proof details differ.

Recall the definition of $g_i$. 
Let $s_1, s_2, \ldots, s_m$ be the items that are inserted and removed from the consistent set 
        object $\id{O}$ during the execution $E_i$, in order of their insertion. 
If we inserted all these items in an empty $\id{O}$ in above order and then removed them one-by-one
        (according to $F$ of our object), $g_i$ is the item that would be removed the last.

Let $P_i$ be the execution prefix of $E_i$ that ends with the insertion of $g_i$.
Let $Q_i$ be an execution interval of $E_i$, starting with an operation immediately after $P_i$
        up to and including the $\lit{remove()}$ operation that returns $g_i$ in $E_i$. 
Finally, let $R_i$ be the execution suffix of $E_i$ consisting of all the operations after $Q_i$.
Define a \emph{trimmed} execution schedule $\tilde{Q}_i$ as $Q_i$ but excluding all (\emph{trimmed}) 
        $\lit{remove()}$ operations that in $E_i$ return items inserted during $P_i$.
In particular, the last removal in $Q_i$ is trimmed and does not occur in $\tilde{Q}_i$
        since it removes $g_i$ that is inserted during $P_i$. 

Observe that while executing $E_i$, every removal that happens during $R_i$ must return an item 
        that was also inserted during $R_i$. 
Otherwise, assume that a $\lit{remove()}$ operation in $R_i$ returns an item $\tilde{g}$ that was 
        inserted during $P_i \cup Q_i$, without the loss of generality before $g_i$. 
When $g_i$ was removed (at the end of $Q_i$), $\tilde{g}$ was already contained in $\id{O}$, 
        so $F(\tilde{g},g_i) = g_i$ holds, contradicting the definition of $g_i$.\footnote{If 
        $\tilde{g}$ was inserted after $g_i$, $F(g_i,\tilde{g}) = g_i$ gives the same result}

Since the number of operations in $P_i$ is strictly smaller than in $E_i$ (as it does not include
        the removal of $g_i$), we use the inductive hypothesis to get an interleaved execution 
        $E_P$ for prefixes $P_i$.
Since $P_i$ also contains at least one operation (insertion of $g_i$), we also use inductive
        hypothesis for execution intervals $\tilde{Q}_i$ and $R_i$ to get interleaved executions
        $E_Q$ and $E_R$.
We start our final iterleaved execution by running $E_p$ from the initial state, and by induction we 
        know the processes do not observe a difference from running $P_i$ 
        in their respective solo executions.
However, after executing $E_P$, the consistent set object $\id{O}$ may not empty and contains 
        all the items that were inserted but not removed during $E_p$.
But we will first show below that after $E_p$, it is possible to indistinguishably execute 
        all operations (trimmed or not) of $Q_i$ of both processes ($i \in \{0,1\}$).
As in the proof of~\propositionref{prp:barrier}, we maintain the invariant that operations 
        in $\tilde{Q}_i$ are executed according to the order in $E_Q$ (with respect to each other). 

Assume contrary and consider the first time $t$ when we cannot indistinguishably schedule
        the next operation without violating the above invariant.
Let us first consider that there is at least one operation yet to be performed from $E_Q$, 
        and the first such operation is without the loss of generality by process $j$. 
Moreover, first assume that the next operation by process $j$ is not a trimmed $\lit{remove()}$.
Using the same reasoning to~\propositionref{prp:barrier}, this \emph{critical operation} must be a 
        $\lit{remove()}$ that based on the state of $\id{O}$ at time $t$ returns some item $s$ 
        instead of item $r$. (for an insertion or indistinguishable removal, we would just run it).
By our assumption all previous operations have been indistinguishable, 
        so $\id{O}$ must also contain item $r$ at time $t$.
Item $r$ was inserted by process $j$ (since it removed $r$ in solo execution $E_j$) and if $s$ was 
        also inserted by process $j$, process $j$ must have observed $r$ and $s$ in the same order
        in $\id{O}$ in its solo execution $E_j$, but in solo execution $r$ was returned, 
        contradicting the consistency property of $\id{O}$.
So, the item $s$ should have been inserted by process $1-j$. 

Assume insertion happened during $\tilde{Q}_{1-j}$.
Since the removal of $r$ was not trimmed from $Q_j$, $r$ must have been inserted during $\tilde{Q}_j$,
        so the corresponding removal that was executed in $E_Q$ observed $s$ and $r$ in the same 
        order, but returned $r$ because of the indistinguishable of $E_Q$, contradicting consistency.
Finally, assume that the insertion of $s$ happened during $P_{1-j}$.
If $F(s,g_{1-j}) = g_{1-j}$ then by $F(r,g_{1-j})=r$ (otherwise the prefix of $E_j$ up to 
        removing $r$ is a barrier) we get that $F(r,s,g_{1-j})=F(s,r,g_{1-j})=r 
        \Rightarrow F(r,s)=F(s,r)=r$ contradicting that $s$ can be removed before $r$.
Otherwise, $F(s,g_{1-j}) = s$ means that $s$ would be removed before $g_{1-j}$, 
        thus the corresponding $\lit{remove()}$ was trimmed from $Q_{1-j}$.
Therefore, process $1-j$ has at least one pending $\lit{remove()}$ from $Q_{1-j}$ 
        (one that returns $s$ in $E_{1-j}$).
We claim that the next removal by process $1-j$ has to be precisely the trimmed 
        $\lit{remove()}$ supposed to return $s$, as otherwise this next removal violates 
        consistency (same items as in $E_{1-j}$ are in $\id{O}$ in the same order).
In this case, we undistinguishably schedule the trimmed operation of process $1-j$ that returns $s$
        and move on.\footnote{If the next operation of process $1-j$ was an insertion, 
        we could have indistinguishably executed it anyway} 
        
Next, consider the case when again there is at least one operation yet to be performed from $E_Q$ 
        by process $j$, but the next operation of the process $j$ is a trimmed $\lit{remove()}$.
Since this trimmed removal is not indistinguishable, say it would return an item $s$ instead of $r$.
Precisely for the same reasons as before, $s$ must have been inserted by process $1-j$ and 
        $F(r,g_{1-j})=r$ still holds because otherwise we have a barrier in $E_j$. 
The case if $s$ was inserted during $P_{1-j}$ works exactly as before: 
        if $F(s,g_{1-j}) = g_{1-j}$, we still get a contradiction $F(s,r) = F(r,s) = r$; 
        if $F(s,g_{1-j}) = s$, then the next removal operation of process $1-j$ exists and must be 
        precisely the trimmed operation supposed to return $s$ in solo execution, 
        which we can indistinguishably execute.  
Now assume $s$ was inserted during $\tilde{Q}_{1-j}$.
If $F(g_{1-j}, s) = g_{1-j}$, using $F(r,g_{1-j}) = r$ we get that $F(r,s) = r$.
By definition of a trimmed operation, $r$ was inserted during $P_j$ 
        and since $E_Q$ is executed after $E_P$, $r$ was inserted in $\id{O}$ before $s$.
Hence, $F(r,s) = r$ implies that it is impossible to return $s$ before returning $r$.
Finally, consider $F(g_{1-j},s) = s$.
But in this case, the next removal operation according to $E_Q$ must be by process $1-j$ (because 
        all previous operations were undistinguishable and $s$, inserted during $E_Q$ 
        by process $1-j$ is to be removed first from $\id{O}$), contradicting our initial assumption.

To complete this portion of the proof, we should consider the case when all operations from 
        $\tilde{Q}_i$ of both processes have been indistinguishably executed, but there are trimmed 
        removal operations left in $Q_0$ and/or $Q_1$ and that we can no longer execute
        indistinguishably.
Since all previous operations have been indistinguishable, $\id{O}$ is not empty, and 
        $\lit{remove()}$ operations are not supposed to return $\id{null}$, because that would
        imply the existence of a mute prefix in a solo execution. 
So, let us assume that the next removal applied to $\id{O}$ would return some element $s$ 
        inserted by process $j$.
If process $j$ has a pending trimmed $\lit{remove()}$, that removal operation must necessarily 
        return $s$ in the solo execution, returning any other element $r$ would violate consistency 
        as $r$ and $s$ are contained in $\id{O}$ in both cases in the same order.
Now assume only process $1-j$ has pending trimmed removals, the next of which is supposed to return 
        item $r$ (based on the solo execution).
First of all, $F(g_{1-j}, g_j) = g_{1-j}$ holds because otherwise we would have a barrier.
Also, since the pending removal is trimmed, $r$ must have been inserted during $E_P$ before $g_{1-j}$ 
        and by definition of $g_{1-j}$, $F(r, g_{1-j}) = r$ is true. 
Assume that $s$ is inserted before $g_j$. 
Then, $F(s, g_{j}) = g_j$ because process $j$ already executed its last trimmed operation that 
        removed $g_j$ while $s$ was already in $\id{O}$.
So, by consistency $F(r,s,g_{1-j},g_j) = F(s,r,g_{1-j},g_j)=r \Rightarrow F(r,s) = F(s,r) = r$
        contradicting that $s$ would be removed before $r$.
If $s$ was inserted after $g_j$, then $F(g_j, s) = s$, and we get 
        $F(r, g_{1-j}, g_j, s) = r \Rightarrow F(r,s) = r$, which is sufficient for contradiction
        because in this case we know for sure that $r$ is inserted in $\id{O}$ before $s$:
        $r$ is inserted during $E_P$ and $s$ is inserted after $g_j$ i.e. during $E_Q$ which we 
        execute strictly after $E_P$.

Finally, after the above process is completed, meaning that all operations from $P_i$ and $Q_i$ for 
        both processes have been executed indistinguishably, we execute the operations of $R_i$ 
        for $i \in \{0,1\}$ according to $E_R$.
We need to show that even though $\id{O}$ was not empty to start with, all return values by removals 
        will still be indistinguishable from the respective solo executions.  
Assume contrary and consider the first removal from $E_R$ executed by process $j$ that returns a 
        item $s$ different from the item $r$ returned in the solo execution $E_j$.
We have shown above that $s$ may not be inserted by operations in $P_j$ or $Q_j$.
If $s$ was inserted by an operation in $R_j$, then in $E_R$ the current removal would have observed 
        $s$ and $r$ in the same order, but there it must return $r$ because by inductive hypothesis,
        $E_R$ is undistinguishable from the corresponding solo execution. 
Now consider the case when $s$ was inserted during $E_P$ or $E_Q$ by process $1-j$.
Since the prefix before $R_j$ can not be mute, 
        there should be at least one item that was inserted by process $j$ in $E_P$ or $E_Q$ 
        but never removed before we started executing $E_R$.
Consider all such items that are in $\id{O}$ right after process $j$ finishes executing 
        $P_j$ and $Q_j$ in isolation, and let $b$ be the item that a $\lit{remove()}$ operation
        on $\id{O}$ would return at that point.
Then we have $F(b,g_{1-j})=b$, because otherwise $b$ would be a barrier.
Recall that all removals in $E_R$ return items also inserted in $E_R$, 
        so $b$ is actually never removed in the solo execution, but $r$ is.
Since $r$ is inserted during $R_j$, after $b$, we conclude that $F(b,r) = r$.         
Finally, we know the last operation of of $Q_{1-j}$ by process $1-j$ running in isolation removes
        $g_{1-j}$ by definition of $Q_{1-j}$, and at the time of that removal, $s$ is contained in 
        $\id{O}$ ($s$ is inserted during $P_{1-j} \cup Q_{1-j}$ and not removed, because
        it was in $\id{O}$ after $E_Q$ during $E_R$ in our indistinguishable interleaved execution).  
Thus, $F(g_{1-j},s)=g_{1-j}$ or $F(s,g_{1-j}) = g_{1-j}$ 
        (based on whether $s$ is inserted in $P_{1-j}$ or $Q_{1-j}$).
Combining above and using consistency we get $F(b,g_{1-j},s,r)=r$ or $F(b,s,g_{1-j},r)=r$ 
        implying $F(s,r) = r$.
In addition we know that $s$ was inserted before $E_R$ started, 
        thus before $r$ was inserted, and hence our removal cannot return $s$ before $r$.
\end{proof}

%% file: algowithregs.tex
\begin{figure}[t]
\DontPrintSemicolon
\hspace{0.5cm}
\begin{minipage}[b]{0.45\linewidth} 
\centering
\setcounter{AlgoLine}{0}
{\small
\begin{algorithm}[H]
\SetKwInput{KwVariables}{Variables}
\KwVariables{
\;$\id{Proposed}[2] = \{\bot\};$ \Indp
\;$\id{O};$}
 \textbf{procedure} $\lit{decide} ( v, id = 0 )$\;
   \Indp
   $\id{Proposed}[0] \gets v$ \;
   \If{$\id{Proposed}[1] = \bot$}
   {
    \textbf{return} $v$ \;
   } 
   \textbf{while}($\id{true}$) \;
   {
    \Indp
        $\id{item} \gets O.\lit{remove()}$ \;
        \If{$\id{item} = \id{W}$}
        {
         \textbf{return} $v$ \;
        } 
        \If{$\id{item} = \id{null}$}
        {
         \textbf{return} $\id{Proposed}[1]$ \;
        } 
        \Indm
   }
   \Indm
   \caption{Pseudo-code for process 0}
\end{algorithm}
}
\end{minipage}
\hspace{0.5cm} 
\begin{minipage}[b]{0.45\linewidth} 
\centering
\setcounter{AlgoLine}{0}
{\small
\begin{algorithm}[H]
 \textbf{procedure} $\lit{decide} ( v, id = 1 )$\;
   \Indp
   $\id{O}.\lit{insert}(\id{W})$ \;
   $\id{Proposed}[1] \gets v$ \;
   \If{$\id{Proposed}[0] \neq \bot$}
   {
    \textbf{return} $Proposed[0]$ \;
   } 
   \textbf{while}($\id{true}$) \;
   {
        \Indp   
        $\id{item} \gets O.\lit{remove()}$ \;
        \If{$\id{item} = \id{W}$}
        {
         \textbf{return} $v$ \;
        } 
        \If{$\id{item} = \id{null}$}
        {
         \textbf{return} $\id{Proposed}[0]$ \;
        } 
        \Indm
   }
   \Indm
   \caption{Pseudo-code for process 1}
\end{algorithm}
}
\end{minipage}
\caption{Two process consensus using a consistent set object $\id{O}$ and registers}
\label{fig:collcons2}
\end{figure}

%% file: algo2obj.tex
\begin{figure}[t]
\DontPrintSemicolon
\hspace{0.5cm}
\begin{minipage}[b]{0.4\linewidth} 
\centering
\setcounter{AlgoLine}{0}
{\small
\begin{algorithm}[H]
\SetKwInput{KwVariables}{Variables}
\KwVariables{
\;$\id{O_0}, \id{O_1};$ \Indp}
 \textbf{procedure} $\lit{decide} ( v, id = 0 )$\;
   \Indp
   $\id{O_0}.\lit{insert}(\{\id{P_0},v\})$ \;
   $\id{O_0}.\lit{insert}(\{\id{Q_0},v\})$ \;

   $a_1 \gets \lit{remLW(\id{O_1})}$\;
   $a_0 \gets \lit{remLW(\id{O_0})}$\;
   \If{$\id{a_0}.\lit{first} = \id{W}$ \textbf{and} $\id{a_1} = null$}
   {
        \textbf{return} $v$ \;
   }
   \Else
   {
        \textbf{return} $a_1.\lit{second}$ \:
   }
   \caption{Pseudo-code for process 0}
\end{algorithm}
}
\end{minipage}
\hspace{0.2cm} 
\begin{minipage}[b]{0.5\linewidth} 
\centering
\setcounter{AlgoLine}{0}
{\small
\begin{algorithm}[H]
 \textbf{procedure} $\lit{remLW} ( O)$\;
  \Indp
   \textbf{while}($\id{true}$) \;
   {
        \Indp
        $\id{t} \gets \id{O}.\lit{remove()}$ \;
    \If{$t = null$}
    {
     \textbf{return} $\id{null}$ \;
    }
        \If{$\id{t}.\lit{first} \in \{\id{P_i},\id{Q_i}\}$}
        {
        $v  = \id{t}.\lit{second}$ \;
        \If{$F(\{P_i, v\}, \{Q_i, v\}) = t$}
        {
         \textbf{return} $\{W, v\}$ \;
         }
        \Else
        {
        \textbf{return} $\{L, v\}$ \;
        }
        } 
        \Indm
   }
  \Indm
%
 \textbf{procedure} $\lit{decide} ( v, id = 1 )$\;
   \Indp
   $\id{O_1}.\lit{insert}(\{\id{P_1},v\})$ \;
   $\id{O_1}.\lit{insert}(\{\id{Q_1},v\})$ \;
   $\id{b_0} \gets \lit{remLW(\id{O_0})}$\;
   $\id{b_1} \gets \lit{remLW(\id{O_1})}$\;
   \If{$\id{b_0}.\lit{first} \neq L$ \textbf{or} $\id{b_1}.\lit{first} = L$}
   {
        \textbf{return} $v$ \;
   }
   \Else
   {
        \textbf{return} $\id{b_0}.\lit{second}$ \:
   }
   \Indm
   \caption{Pseudo-code for process 1}
\end{algorithm}
}
\end{minipage}
\caption{Two process consensus using two consistent sets objects $\id{O_0}$ and $\id{O_1}$}
\label{fig:2objcons2}
\end{figure}

%% file: imposs.tex
\section{Unbounded Number of Objects}
\begin{theorem}
\label{thm:nocntbound}
It is impossible to implement an isolation-bounded test-and-set object for an unbounded number of 
        processes using any number of (possibly infinitely many) empty queues (or empty stacks).
\end{theorem}
\begin{proof}
Let us assume contrary and consider an isolation-bounded algorithm 
        that implements test-and-set for an unbounded number of processes with initially empty queues.
Because of isolation-boundedness, any process that runs in isolation from the initial state 
        can take at most a fixed number of steps, say $M$,
        each being an $\lit{insert(\id{item})}$ or $\lit{remove()}$ operation on one of the queues,
        before returning $1$.

Associate to each process $p$ the ordered list $s_q$ of the $M$ steps 
        it would take if it ran in isolation. 
We call this quantity the \emph{signature} of $p$.
Suppose each queue is touched by finitely many signatures. Let $Q_1$ be any queue which is touched,
        say by process $p$. Then $p$'s signature touches at most $M$ queues, 
        call them $Q_1, \ldots, Q_M$. 
At most finitely many other processes can touch these same queues, so there must be a process $q$ 
        whose signature does not touch any of the $Q_i$. Running $p$ then $q$ gives us an immediate 
        contradiction, since their actions on the queues they touch do not interact at all, and thus 
        they cannot distinguish between running together and running in isolation, 
        and must both return $1$.
        
Thus we can assume that there exists a queue $Q_1$ such that an operation on this queue occurs 
        in infinitely many signatures. 
Let $\mathcal{P}_1$ denote the set of processes whose signatures contain an operation on $Q_1$.
Next, if there is a queue $Q_2$ such that an operation on it occurs in infinitely many 
        signatures from $\mathcal{P}_1$, we consider this infinite subset 
        $\mathcal{P}_2 \subseteq \mathcal{P}_1$. 
Inductively, we build sets 
        $\mathcal{P}_i \subseteq \mathcal{P}_{i - 1} \subseteq \ldots \subseteq \mathcal{P}_1$ and 
        choose queues $Q_i$, until the process terminates.
This can only happen at most $M$ times, since the members of $\mathcal{P}_M$ (if they exist) must in 
        isolation perform the maximum number of allowed operations (i.e. $M$ operations), 
        namely on the queues $Q_1, \ldots, Q_M$.
Thus, we end up with an infinite set of signatures $\mathcal{P}_m$ ($m \leq M$), such that 
        each of the signatures contains an operation on each $Q_j$ ($1 \leq j \leq m$), 
        and for every other queue, an operation on it is contained only in a finite number of 
        signatures from processes in $P_m$. We let $\mathcal{Q} = \{Q_1, \ldots, Q_m\}$.

We can now find an infinite subset $\mathcal{P} \subseteq \mathcal{P}_m$, such that if two processes 
        from $\mathcal{P}$ have signatures which involve operations on a shared queue, this
        queue has to be one of our selected queues $\mathcal{Q}$. 
We do so inductively: choose $p_1 \in \mathcal{P}_m$ arbitrarily. 
This process's signature touches at most $M - 1$ queues not in $\mathcal{Q}$.
Moreover, finitely many other processes in $\mathcal{P}_m$ have signatures which touch these queues 
        by the construction of $\mathcal{P}_m$. 
Thus we can choose a $p_2 \in \mathcal{P}_m$ which does not touch any of these queues, and then 
        we recurse to find $p_i$ for all $i$, and we let $\mathcal{P} = \{p_i\}_{i = 1}^\infty$. 
It is straightforward to verify that this set has the desired property.

Let us now focus on the processes in $\mathcal{P}$ and consider only the operations they perform on 
        queues $\mathcal{Q}$.
Clearly, each process performs at most $M$ such operations when run in isolation.
Each operation is either $\lit{insert(\id{item})}$ or $\lit{remove()}$ on some $Q_j$,
        thus there are $2m$ different types of operations.
There are only finitely many different possibilities to order at most $M$ operations of
        $2m$ different types, and infinitely many processes in $\mathcal{P}$, thus by the
        pigeon-hole principle, we can find two processes $p, q \in \mathcal{P}$, such that their 
        signatures both involve the same operations on the same queues in $\mathcal{Q}$ in exactly 
        the same order. 
Moreover, they may perform actions on queues not in $\mathcal{Q}$, but by the construction of 
        $\mathcal{P}$, the sets of queues they touch outside of $\mathcal{Q}$ are disjoint.

Let us execute $p$ and $q$ in the following ``lock-step'' fashion: 
        we let $p$ take steps until the first operation on some $Q_j$, then 
        we let $q$ take its steps until it performs the same type of operation on the same $Q_j$, 
        etc, until they both finish.
At any point in the execution when $q$ has just taken a step, 
        we claim that the following invariant holds:
        none of the processes have observed a difference from their solo executions,
        and each queue $Q_j$ contains items that $p$ inserted and items that $q$ inserted, 
        interleaved one-by-one.
Moreover, if we only consider the items inserted by one of the processes, say $p$, they are 
        the same items and in the same order as in the solo execution of $p$. 

$p$ and $q$ could only observe a difference after a $\lit{remove()}$ call on one of the 
        queues $Q_j$, because other queues are accessed by only one process.
Now, the invariant holds initially, and if the next operation on some $Q_j$ is insertion 
        (necessarily the same queue for both processes, but they may insert different items),
        we let $p$ insert, then $q$ insert, so the invariant holds afterwards.
If it is a removal from some $Q_j$ for both processes, then since the items of $p$ and $q$ are 
        interleaved but consistent with respective solo executions, first removal by $p$ will 
        return the item $p$ previously inserted (or $\id{null}$) and does not observe a difference,
        then $q$ does the same with its item.

Thus, we are able to execute $p$ and $q$, both of which cannot distinguish the execution from a solo
        execution and return $1$ contradicting the correctness of the test-and-set implementation.

A very similar argument works for the stack, except when running processes in lock-step, if the 
        operation is a $\lit{remove()}$, we should reverse the order and let $q$ execute first.
\end{proof}
\noindent On the other hand, if we have registers available implementing test-and-set 
        becomes possible.
\begin{theorem}
\label{thm:dantourney}
It is possible to implement an isolation-bounded test-and-set object 
        for an unbounded number of processes using infinitely many 
        consistent set objects (in any initial configuration) and read-write registers.
\end{theorem}
\begin{proof}
The adaptive tournament tree from~\cite{AAGGG10} is an algorithm that implements isolation-bounded
        test-and-set for an arbitrary number of concurrent processes.\footnote{We consider 
        non-randomized version of the construction.}
It requires registers and a black-box test-and-set primitive for two processes.
Using~\lemmaref{lem:2obj}, we can do test-and-set for two processes with just two consistent set 
        objects initialized with a finite number of arbitrary items in an arbitrary order
        (or with one object and registers, per~\lemmaref{lem:collcons2}).
This two process test-and-set object can be directly plugged into the~\cite{AAGGG10} construction
        as the building block.
The other crucial building block is a splitter object~\cite{AM94}, 
        which is easily consructed using registers.
The algorithm is isolation-bounded, since any process running in isolation from the 
        initial state stops in the first splitter and participates only in a few 
        two-process test-and-sets.
\end{proof}
\begin{corollary}
\label{clr:pwrreg}
It is impossible to implement a read-write register in an isolation-bounded way
        using any number of (possibly infinitely many) empty queues (stacks).
\end{corollary}
\begin{proof}
Assume contrary.
Then we can use the same algorithm as in~\theoremref{thm:dantourney} to implement a test-and-set 
        object for an unbounded number of processes, except we replace each register in the 
        construction with an isolation-bounded register implementation out of empty queues.
The resulting test-and-set construction would then only use empty queues and would be 
        isolation-bounded, because both the original implementation and the new register 
        implementation are isolation-bounded.
In fact, if the constant bounds on the number of steps are $c_1$ and $c_2$,
        the bound for the new construction would be $c_1 c_2$.
Such a construction, however, contradicts~\theoremref{thm:nocntbound}.  
\end{proof}
\begin{corollary}
\label{clr:initstate}
It is impossible to implement a queue (a stack) containing one element in its initial state
        using any number of (possibly infinitely many) empty queues (stacks) 
        in an isolation-bounded way.
\end{corollary}
\begin{proof}
By~\lemmaref{lem:collinftas}, a single consistent set object initialized in a lucky state
        can implement a wait-free test-and-set object for unbounded number of processes.
A queue is a consistent set object and a state with a single item is a lucky state.
By inspection, the test-and-set algorithm from~\lemmaref{lem:collinftas} using a queue with 
        a single element is isolation-bounded (an initial isolated run involves just one removal).
Therefore, being able to implement a queue with a single item would immediately allow implementing 
        an isolation-bounded test-and-set object for an unbounded number of processes, 
        which by~\theoremref{thm:nocntbound} is impossible using any number of empty queues. 
\end{proof}